## Reliability in the face of complexity – the challenge of high-end scientific computing

Gary Ferland

*Physics Department, The University of Kentucky, Lexington, KY 40506*

**Abstract.** This conference focused on photoionized plasmas and the tools necessary to understand them. One of the major activities was a series of sessions that brought together the developers of several plasma emission codes. The purpose was to identify and resolve sources of differences between the various programs. Very often these differences were caused by bugs. This contribution outlines some practical aspects of developing bug-free code and is based on my experience creating the plasma simulation code Cloudy.

Computers present nothing less than a new way to understand the universe. We can do any numerical experiment that we can imagine. But machines are growing faster far more quickly than people are getting smarter. More powerful machines allow more complete and complex simulations of physical processes but also offer a far greater opportunity for errors. Harnessing the power of faster computers in a reliable way will present an ever greater intellectual challenge. Modern standards of quality control, and the associated code development techniques they require, can play a major role in developing reliable astrophysical simulations.

## 1 Introduction

Computers, and the numerical simulations they make possible, are nothing short of a new way of understanding the universe. These simulations allow us to observe what happens when different hypothetical situations occur, and so gain insight into what may happen in nature. Of course, none of this is meaningful if the simulations do not obtain the correct answer. That brings up the subject of this paper, the very difficult question of quality assurance; how to get the right answer in a large-scale numerical calculation. In the case of plasma codes it is important to make correct predictions because these programs are tools that are used to understand astronomical observations, and so nature itself.

Creating a numerical simulation should, at first thought, be simple. With a powerful computer we can do anything we can imagine. Producing the program itself should be simple because, like a poem, a program is composed only of thoughts. This notion is deceptive – it certainly is relatively easy to create a code that does not crash, but can we create one that we *know* gets the right answer? It is the nature of today's computer languages that the machine does precisely what we tell it to do. It will get the right answer as long as the directions we give it are perfect. Put another way, a modern plasma code, roughly as long as *Gone with the Wind*, could obtain worthless results with an error that is equivalent to a single misplaced comma in the novel.

We humans are very good at creativity, and imagining things that do not exist, but we are not very good at producing complete perfection – creating the precise instructions that a computer needs. In the future, high-level languages will be the norm and it will be possible to solve a problem by merely describing it. But for now a large physics code must use an algorithmic language like C or Fortran. How can we create a very large code in a reliable way, detect any mistakes that are present, and be confident that it obtains the right answers? This problem is exacerbated by the fact that the correct answer is usually not known beforehand – how can we tell whether errors are present when we don't know what the answer is supposed to be?

Humans are also naturally optimistic – if we were not we would not try to create a large computer program. Somehow the right thing will happen. Unfortunately, we can't be confident that the code gets the right answer just because it does not crash. Validation is a deep process that involves coding methods that introduce as few errors as possible, the rapid and automatic detection of any errors that are introduced, and extensive testing.

One of the best ways to validate large codes is to compare results of several independent calculations. This conference is part of a series that was originated by Daniel Péquignot in 1985. The idea is to make detailed comparisons of results from independent codes. This will reveal differences in numerical algorithms, any variations in the assumed atomic physics cross sections and rate coefficients, and the presence of bugs. The results of this comparison will be twofold; first, a set of benchmarks that represent the current state of the art, and second, a set of plasma codes with better approximations and atomic data, and with fewer bugs.

This paper is a summary of the tricks and methods I have learned over the past 25 years for writing error-free code. (Actually, code that has several orders of magnitude fewer errors than before, but unfortunately, even this is not good enough). Although these methods do take more time in the short run, in the long term much less time will be spent chasing bugs, and incorrect results are less likely to be published. The discussion is heavily influenced by two books, *The Mythical Man-Month* (Brooks 1995) and *Writing Solid Code* (Maguire 1993).

## 2 The surgical team

Integrity of design with a clear central vision of the goal is the most important need in developing a large code. Brooks (1995) uses the example of a medieval cathedral - they were often built over hundreds of years by thousands of masons, but the final product still managed to carry out the design and intent of the original architects. If each generation of mason had gone off in their own direction the result would have been a mess. This required both a clear original intent and a commitment by the builders to carry out this design. For software development to be successful this same integrity of design must be achieved.

The "surgical team" (Brooks 1995) is the approach I use in developing Cloudy, and (I think) the one best suited for developing codes for astronomical simulations. By industrial standards these codes are "medium-sized" projects, with sizes of roughly a few hundred thousands of lines of code. Far more structured approaches are needed for large projects, with millions of lines of code, while less formal methods can be used for small projects.

The surgical team metaphor visualizes a single skilled surgeon with complete responsibility for the project but with (perhaps) a few helpers. The surgeon guides the entire operation, has a clear vision of the goals, and is the person who does the most important work. There may be many assistants in a surgical team, but the surgeon is the one holding the scalpel. This approach is the most efficient way to develop medium sized codes, where it is still possible for a single individual to have a detailed understanding of the whole design, and where implementing that central concept is the goal.

I know one method that does not work – "the high priest – hapless postdoc" approach. Here a senior investigator has a general idea of the long-term goal, perhaps with an understanding of the associated physics, but little detailed knowledge of the actual code. Details of the implementation are left to a "project postdoc", someone hired on a limited term contract to come in and execute the design. Often the project will consist of extensive modifications to an existing code, perhaps written elsewhere, or by an earlier postdoc. This approach generally produces a program that is understood by no one. The basic problems are that the intricacies and intercouplings that exist between various components will not be obvious to a casual observer, there is no persistent knowledge about parts of the code written by previous postdocs, and there will not be an overall grand picture of the design.

## 3 Some practical tricks

There are a series of steps one can take in developing software, each of which can reduce bugs by something like an order of magnitude. Unfortunately even that is not good enough – a single bug can be fatal. Cross checks between independent codes such as those done in this conference remain vital.

These steps fall into three general categories. First are techniques that are designed to reduce the number of new errors that are introduced. This mainly involves methods of coding, and checking code as it is produced. The second is to create the infrastructure within the code to catch bugs when they are introduced. The last is to fully test every part of the code every time anything is changed.

We want a compiler so smart that it will tell us if we make a logical error (Maguire 1993). Since software vendors cannot provide this, we must accomplish it by creating the infrastructure within the program. All of these steps are designed to do just that – when problems occur the program must be the first to report them.

Good practices demand that bugs be caught as soon as possible after they are introduced. This is basically the "First Rule of Holes": if you find yourself so deep in a hole that you can't get out, first of all, stop digging. By detecting bugs soon after they are introduced we can discover whether we have dug ourselves into a hole. Then we can stop making them worse.

Most of these methods are standard techniques in the real world of commercial software development. But most of us are self-trained amateur programmers and may be unaware of what is out there.

The next section outlines the first of the three steps – safe coding methods. Remaining sections outline the creation of a self-aware code and the automatic validation of all predictions.

### 3.1 Safe coding methods

***Failsafe tests***   These are described in most beginning computer science texts. The basic rule is to never assume that the impossible cannot happen, and always have a plan to detect it when it occurs.   The following table gives one example (among many possibilities) of safe and unsafe ways to conduct a simple test.

| Unsafe | Safe |
| --- | --- |
| ```
if key = 1
    do the key=1 job
else
    do the key=2 job
``` | ```
if key = 1
    do the key=1 job
else if key = 2
    do the key=2 job
else
    declare insanity
``` |

Suppose the variable "`key`" must be either 1 or 2, and what we do next is determined by this value.   The left column of the table shows an example of unsafe code.   The author knew that `key` had to be either 1 or 2, so only the first value was checked.   This is unsafe because memory errors (exceeding an array's bounds, for example) could give `key` an impossible value, or there may be logical paths where `key` is undefined and so perhaps equal to zero.   The only symptom would be that the `key=2` job was always done since it is the default.   It would take some effort to discover why this happened.   The right column shows the safe way to make the test.   If something goes wrong and `key` is either not set or is disturbed, the problem will be detected automatically the first time the code is executed.   (Zero should not be a possible value, since zero may occur if `key` is never set at all).

***A consistent style convention***   It should be possible to guess a function's purpose or a variable's type by examining its name.   FORTRAN IV had some of this – a variable starting with certain letters was known to be of a certain type. The so-called Hungarian convention is the most popular of the current naming conventions (Simonyi 1977), but I use a mix of the old Fortran styles and the new convention.   For instance, character strings begin with "ch", logical variables with "lg", and so on.   Routines have names with the format "noun_verb" where "verb" is the job the routine does, and "noun" is the thing that receives action.   Routine "HeatSum" sums all heating agents and "HydroCreate" creates space and data for the hydrogenic iso-electronic sequence.

These style conventions extend beyond names to such details as the location of braces around loops and indentations that show the logical flow within the code.   All of these are important because they establish "handwriting" across the code, a style that can be easily understood in the context of the rest of the program.   This makes the code simpler to understand and so easier for the maintainer to detect errors.

***The principle of least astonishment***   It should be totally clear what code does, with no surprises or bizarre twists.   Clarity is far more important than fast or clever because simple code can be more easily debugged, maintained, and understood.   Clear coding methods are closely related to safe coding methods, since this is part of the methodology for producing safe code in the first place.

For instance, a quantity such as the ratio of densities of two ion densities should have a name that indicates this (for instance, how about CIV_2_HeII?). When this variable is set, it should clearly be equated to the ratio of the



densities of these two species. Clever code that saves a division while obfuscating the intention will cost time in the long run and is likely to hide bugs. Simple and clear is best.

**Self-documenting code** The idea here is to be able to have a document extraction program like Perl automatically create documentation for the code. This requires a uniform style for introducing comments. For instance, all tags could begin with a unique string that can be located with a pattern searcher, and be followed by information stored in a standard format.

The atomic data used by Cloudy are one example. A code like Cloudy can only exist because of the work of many dedicated atomic physicists, and this work must be properly cited. All atomic data have an associated comment with a string denoting the species (i.e., $H^0$, $C^{+6}$, $H_2$), the type of data (transition probability, collision strength, recombination rate coefficient, etc) and the original journal reference. A Perl script can then automatically generate a complete atomic data bibliography giving the type of data and its origins.

Changes to the code are similarly documented, this time with a comment containing the date and nature of the change. This makes it easy to identify all changes that occurred after a certain date, or on the day that something broke. (Source code control systems also can do this, but I have not found them worth the effort in a code as small as Cloudy).

**Code review** It is notoriously difficult to find typos in your own papers. The problem is that the eye sees what the mind knows – we interpolate over our mistakes. Two good ways to proofread a paper are to either have someone else look at it, or set it aside and read it after a day or two.

Code review is the same concept. Many corporations have pairs programmers review one another's code as part of their duties. A second person can more quickly and easily spot any mistakes made by the first. This is a much more cost efficient approach than finding problems later through testing and debugging. This has even led to the concept of "extreme programming", in which two people actually share a common keyboard as code is written.

Since projects in astronomy are usually too small to allocate two programmers per keyboard, other methods must be devised. Willing collaborators can help a lot. Also, the Open Source movement has created an ethic where code review has become normal – third parties are able to see the code for themselves and discover how a result was obtained. Cloudy has always been openly available, in keeping with the NSF's policies on public access to project results. This has proven to be an extremely valuable way to validate code, but one that happens on a relatively long timescale –user feedback on bugs in new code may not come in for months or a year.

Although having another person review new code is the best strategy, there are often circumstances where this is not possible. Many of the same benefits are obtained by reviewing new code a day or two after it was written. But the time spent in this review will be small compared to the time that would be spent in debugging if it were not done.

**Single step through new code with the debugger** This is a surprisingly efficient way to find problems and is a form of code review. All modern IDEs (Integrated Development Environments) include a graphical debugger. New code should be checked by setting a breakpoint at the start of new material and observing the logic flow, step by step, as the code is executed. This seems to make a major difference in how you see the newly written code and its bugs.



***Use all the help you can get***   Set the compiler's warning flags to be as finicky as possible. Use every version of lint on the system. Invest in commercial bug finders such as *Insure* or *Purify*. All of these initially cost some time and effort but usually result in better and safer coding practices and final results.

## 3.2 An autonomous and self-aware program

We would like to have a compiler that is so smart that it would tell us if we made a logical error (Maguire 1993). These don't exist, but we can effectively build one by producing a program that is smart enough to detect errors when they occur. The code must be autonomous and self-aware, something that can only happen if we create the appropriate infrastructure. The general idea is have multiple and redundant verifications that everything that is happening makes sense.

***Validate user input.***   All input to the code must be checked and validated by the code as it is read in, since this is the one aspect of the simulation that is totally out of control. This must be done at all times for all input. Unexpected parameters, or parameters that do not make physical sense when used together, should be pointed out.

***Assert the obvious.***   Each routine that uses quantities should make sure that they are physically correct. Are the electron density and temperature positive? Are the atomic transition probabilities still correct? Is the wind velocity less than $c$? This can be done with hand coded logical tests in Fortran. The C language provides the `assert` macro – it has no effect if its argument is a valid test, but the program will throw an exception if the test fails. The assert tests do not slow down optimized code since they become empty statements when the proper compiler flags are set.

***Multiple validations of important results and routines***.   The ability of important routines to generate correct results should be verified at startup. For example, at the start of every new Cloudy simulation the numerical quadrature routine is used to integrate the sin function from 0 to $2\pi$, the matrix inversion routine solves a trivial 2x2 system of equations, and several transition probabilities are checked. Some insidious bugs are first detected when such trivial test calculations fail unexpectedly.

Passing known quantities through the code's infrastructure can also catch problems. Cloudy predicts the intensity of a dummy optically thin emission line with an emissivity per unit volume of unity. This "line" is treated like all others – it is passed through the routines that enter lines into the emission-line arrays and those that integrate over the computed structure to obtain the total line luminosity. At the end of the calculation the "luminosity" of this dummy line must be equal to the volume of the emitting region.

In other cases the same quantity can be calculated using different algorithms, and the results compared. For instance, at the time of this writing Cloudy has two completely independent model helium atoms – the original one, dating back to the mid 1980's, and a new one, now under development, which does the entire helium-like isoelectronic series. Predictions of the two can be compared for atomic helium and any differences must be carefully understood. Significant improvements to the physics are incorporated into both versions so that they stay relatively similar. The two versions thus check one another, although the older version will be removed once development is complete. This insures that



one functioning helium atom is always available for reference, and provides some confidence that both are correct.

***Write code to automatically detect newly found bugs before fixing them*** We want a code that automatically detects mistakes. The mistakes we have made in the past are the best predictor of the mistakes we will make in the future. When a bug is discovered the first thing to do is to write code to automatically detect the problem. Then run the code to check that the error is caught and the problem explained. Only then should it be fixed. This means that the next time this or a similar mistake is made (which will probably happen since it already has), the error will automatically be caught and a diagnosis printed.

***Bugs come in groups of three*** This sounds like the type of old programmer's tale that can't be true. My experience is that this is more than an urban legend – bugs do come in clusters (but not necessarily three!). When a mistake was introduced it was probably because of a mistaken mindset or misunderstanding of the consequences of some action. Other very similar mistakes were probably introduced at about the same time. When one bug is found, look for similar ones in associated code.

### 3.3 Validation of results

The *entire* code must be *completely* retested every time *anything* is changed. This is because of the many subtle interrelationships between apparently unrelated physical processes and code variables, and because a bug in one part of the code may only be manifested in other apparently unrelated parts.

Bugs must also be caught as soon as possible after they are introduced (the First Rule of Holes). This is the easiest time to observe their effects (predictions probably changed), they have not had time to become part of the code's infrastructure, and you still remember the changes that have been recently made.

Here in Lexington, Cloudy is recompiled and completely exercised *every single night*. Errors are automatically detected by the Perl script that conducts the tests, and email is sent to announce the result of the tests when they end. The first task when arriving for work in the morning is to fix whatever broke the day before.

The capability for automatic testing has to be designed into the code. I developed a series of *assert* commands for Cloudy, analogous to the C assert macro, but included as a command in the code's input stream. These tell the code what answers to expect for various quantities. These might be a particular simulation's H$\beta$ luminosity, the average temperature of the He$^+$ region, or the CO column density. These expected values can be based on analytical predictions or the answer the code has obtained previously.

These assert commands are included in a test suite comprised of over a hundred simulations that span the physical conditions that Cloudy can model. Each test highlights a different aspect of the nebular physics. The goal is to have every one of the code's features exercised somewhere in this test suite. Cloudy prints a message with a standard string if predicted quantities are significantly different from their expected values, or if numerical stability problems occur. The results are verified by searching for the standard string that denotes a problem, and email is sent to announce the results. The result is that new bugs are usually detected within 24 hours of their introduction, making



it both easy to fix them ("how did yesterday's change in the CO rotation levels break the Compton equilibrium of $10^6$ K gas?") and minimizes their affects.

Once automatic validation is in place it is a simple matter to test the code on many different computers. This is a surprisingly efficient way to find problems. Different compilers will implement different types of "sanity checks" – one compiler may flag suspicious code that others do not. Answers that change from one machine to another probably point to code that is either ambiguous or on the ragged edge of numerical stability.

## 4 Conclusions

Computers are a new way to understand the universe. We can simulate anything we can imagine, but we must do it perfectly. It is notoriously difficult to change the mathematical description of what we intend to do into the explicit instructions that a computer needs. This paper has not discussed any numerical algorithms or physical results at all, but rather quality assurance and the day to day aspects of developing a code that will probably get the right answer. Quality assurance has to be given the same priority as the astrophysics – the alternative is a faulty simulation and a misleading view of what is happening in front of our telescopes.

Most developers of large theory codes are self-trained amateur programmers. I could teach myself how to use computers largely because machines were so feeble when I began. I became better at using them as they became more capable of solving complex and difficult problems.

But today we expect a graduate student to become self-educated in programming techniques, write a code that may take hours running on a 128-way machine, *and get the right answer*. It strikes me as naïve to expect a student to do this by trial, error, and self-study. We do not expect them to become self-educated in analytical mathematical techniques such as partial differential equations – extensive under and post graduate education is provided.

Should preparation for a career in astrophysics include formal instruction in both numerical methods *and quality assurance*? How important is it to get the right answer? I think that incorporating modern software development methods into the curriculum would help both a student's astrophysics and job prospects should they not remain in astronomy.

***Acknowledgements***; My attempts at keeping bugs out of Cloudy have been supported by the NSF (AST-0071180), NASA (NAG5-8212), and by the Center for Computational Sciences at The University of Kentucky.